\documentclass[prl,twocolumn,superscriptaddress]{revtex4}
\usepackage{graphicx}
\usepackage{amssymb}
\usepackage{amsmath}
\usepackage{bm}
\usepackage{color}

\begin{document}

\title{The Design for a Nanoscale Single-Photon Spin Splitter}

\author{G. Li}
\affiliation{School of Physics and Astronomy, University of Southampton, SO17 1BJ, Southampton, United Kingdom}
\affiliation{School of Physics and Astronomy, Monash University, Victoria 3800, Australia}

\author{A. S. Sheremet}
\affiliation{Russian Quantum Center, Novaya str. 100, 143025 Skolkovo, Moscow Region, Russia}

\author{R. Ge}
\affiliation{Division of Physics and Applied Physics, Nanyang Technological University, Singapore}

\author{T. C. H. Liew}
\affiliation{Division of Physics and Applied Physics, Nanyang Technological University, Singapore}

\author{A. V. Kavokin}
\affiliation{School of Physics and Astronomy, University of Southampton, SO17 1BJ, Southampton, United Kingdom}
\affiliation{Russian Quantum Center, Novaya str. 100, 143025 Skolkovo, Moscow Region, Russia}
\affiliation{SPIN-CNR, Viale del Politecnico 1, I-00133 Rome, Italy}
\affiliation{Spin Optics Laboratory, Saint Petersburg State University, 1 Ulianovskaya, 198504, Saint Petersburg, Russia}

\begin{abstract}
We propose using the effective spin-orbit interaction of light in Bragg-modulated cylindrical waveguides for the efficient separation of spin-up and spin-down photons emitted by a single photon emitter. Due to the spin and directional dependence of photonic stopbands in the waveguides, spin-up (down) photon propagation in the negative (positive) direction along the waveguide axis is blocked while the same photon freely propagates in the opposite direction.
\end{abstract}
\maketitle

\emph{Introduction.} The development of nanophotonic devices brings an emergent research field of chiral quantum optics (CQO) \cite{LodahlNature17} in microscopic waveguides. When light is confined strongly in the transverse direction, its electromagnetic field oscillates along both transverse and longitudinal directions, resulting in a rotating electric field oriented perpendicular to its propagation direction and forming the transverse spin \cite{AielloNatPho15,BliokhPhysRep15,BliokhScience15}.
The transverse spin is locked to momentum because its component flips sign with the inversion of propagation direction.
When an emitter is embedded inside the waveguide, the light absorption and emission depend on the local distribution of the momentum-locked transverse spin, resulting in the CQO \cite{LuxmoorePRL13,JungePRL13}.

The transverse spin is an exemplification of a more general concept of spin-orbit interaction (SOI) of light, which arises due to the vectorial nature of the light field encountering wavelength-scale structures \cite{BliokhNatPho15}. Besides the transverse spin, the SOI of light results in phenomena such as the spin-Hall effect of light \cite{OnodaPRL14,BliokhPRL06,HostenScience08,AielloOptLett08,BliokhNatPho08,GorodetskiPRL12,BliokhJOpt13} and spin-to-orbital angular momentum conversion \cite{DogariuOptExp06,ZhaoPRL07,BliokhOptExp11}.
Currently, many CQO designs rely heavily on chiral atom coupling \cite{JungePRL13,MitschNatPho14,ShomroniScience14}, making them hard to implement in integrated optical circuits \cite{IOCbook}. To circumvent this constraint, here we realize a CQO design serving as a fully optical single-photon spin splitter by exploiting the SOI of light in cylindrical waveguides \cite{LearyPRA09,LearyPRA14,VitulloPRL17} without introducing any light-matter interaction.
To achieve this, we need to combine effects from the SOI of light and the band gap structure of a photonic crystal.

When light is passing through a periodically modulated dielectric structure,
the electric field tends to concentrate around the high refractive index regions \cite{JoannopoulosBook08}; and a finite amount of energy would be required to change the electric field to the reverse distribution. Therefore, for a certain range of frequency there will be no propagating mode, i.e. the appearance of a photonic band gap \cite{JoannopoulosBook08}. Recent developments in fabrication have allowed embedding an ultra-long Bragg modulation (up to $1$ meter) in a waveguide \cite{GagneOptExp14}. Hence, a question will arise as to how the photonic band gap structure under the influence of transverse confinement is modified by the SOI of light.
In this letter, we will try to answer this question by analyzing the SOI in the presence of a weak Bragg modulation in a cylindrical waveguide. We present a systematic method in dealing with transversely confined periodic structures by making clear the distinction between the Floquet exponent (i.e. the Bloch wave vector) \cite{KomlenkoBook05,FloquetTheory} and the variable separation constant \cite{MyintBook07}, clarifying some misunderstandings found in the literature \cite{CarreteroOptExp06,PereyraHindawi17}.
The SOI will lead to splitting of the propagation constants between $\sigma_+$ and $\sigma_-$ components in the helicity basis \cite{LearyPRA09,LearyPRA14,VitulloPRL17}. As a result, the photonic band gap will split accordingly and lead to spin-locked propagation modes protected by the photonic band gap structure, forming the foundation of our design of a single-photon spin splitter.

We will first introduce a general scheme for the dispersion calculation of Bragg mirrors under the influence of transverse confinements, and after that, the effect of SOI correction on the dispersion and band gap structures.

\begin{figure}
  \centering
  \includegraphics[width=8cm]{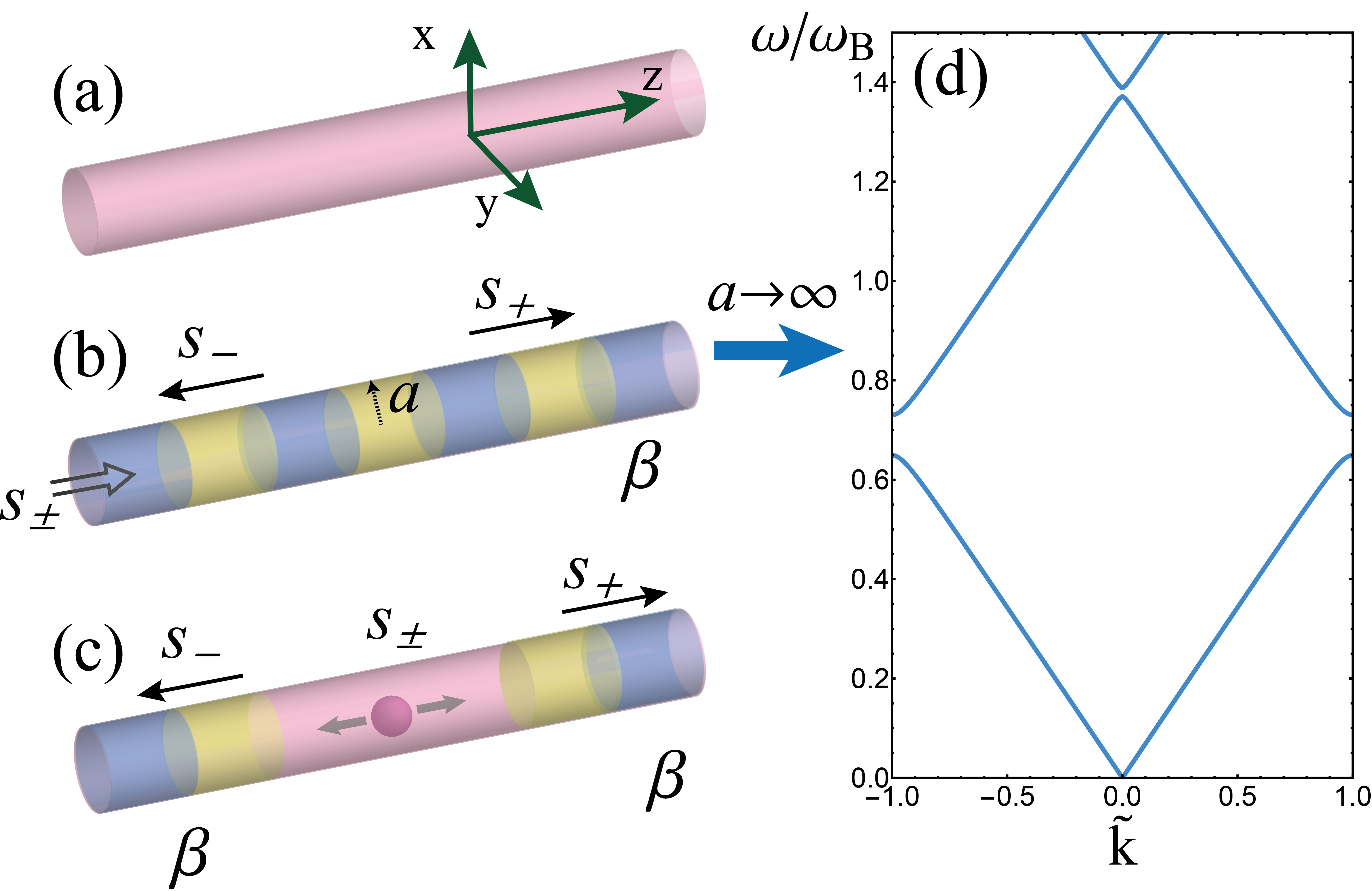}
  \caption{Illustrations of dielectric constant modulations in cylindrical waveguides.  (a) A cylindrical waveguide with homogeneous dielectric constant along $z$ direction. (b) A cylindrical waveguide with weak Bragg grating along the $z$ direction. The SOI effect results in spin-and-direction locked propagation for two polarization components $s_\pm$ in laboratory coordinates. (c) A single photon emitter is sandwiched between two Bragg gratings, serving as the fully optical single-photon spin splitter. (d) Dispersion relation for the TE/TM mode when the radius $a$ of configuration (b) becomes infinite, retrieving the dispersion of an 1D photonic crystal slab. Parameters: $\varepsilon_1=1.12, \varepsilon_2=0.5, \beta=4.2\,\text{$\mu$m$^{-1}$}, \text{and}\, \omega_B=c \beta=1.261\times10^3$ THz.}\label{Waveguide_illustration}
\end{figure}

\emph{Floquet theory and dispersion relation.} 
Let's consider an infinite cylindrical waveguide with weak Bragg grating along the $z$ direction [Fig.~\ref{Waveguide_illustration}(a,b)]. Its relative permittivity has spatial variation and can be written as:
\begin{equation}\label{dielectric_distribution}
  \varepsilon_r(r,z)=1+H(a-r)[\varepsilon_1-\varepsilon_2 \cos(2\beta z)],
\end{equation}
where $H(r)$ is the Heaviside step function, $a$ is the waveguide radius, and $\beta$ represents the spatial frequency of the Bragg grating, $\varepsilon_{1}$ and $\varepsilon_{2}$ are two constants with $\varepsilon_2<1$, and $r, \phi, z$ are cylindrical coordinate variables.

In what follows we shall calculate eigenmodes of the waveguide for corresponding $\varepsilon_r(r,z)$.
By setting the time dependence of the electric field $\mathbf{E}=(E_r, E_\phi, E_z)$ as $e^{-i\omega t}$, $\mathbf{E}$ satisfies a vector Helmholtz equation \cite{JacksonBook}.
In cylindrical coordinate, the vector Laplacian operator can be separated into transverse and longitudinal components (see the Supplemental Material); and  specifically, the $z$ component of the electric field $E_z$ fulfills the equation
\begin{equation}\label{Helmoltz_equation_Ez}
  \nabla^2 E_z+\frac{\omega^2}{c^2} \varepsilon_r(r,z) E_z=0,
\end{equation}
where $c$ is the speed of light and the relative permeability $\mu_r=1$ is assumed.
Separating the coordinate variable dependency as $E_z(r,\phi,z)=R(r)\Theta(\phi)Z(z)$, for $r<a$, we have
\begin{equation}\label{RTZ_1}
\begin{split}
   \frac{r^2}{R}\frac{d^2 R}{d r^2}+\frac{r}{R}\frac{d R}{dr} &+\frac{1}{\Theta}\frac{d^2 \Theta}{d \phi^2}+\frac{r^2}{Z}\frac{d^2 Z}{d z^2}\\
   &+r^2\,\frac{\omega^2}{c^2}\left[1+\varepsilon_1-\varepsilon_2 \cos(2\beta z) \right]=0.
\end{split}
\end{equation}
The solution for $\Theta(\phi)$ satisfies $\frac{1}{\Theta}\frac{d^2 \Theta}{d \phi^2}=-m^2$, with $m$ the quantum number of the angular momentum operator $\hat{l}_z=-i\partial_\phi$ \cite{LearyPRA14}. The $Z(z)$ function satisfies
 \begin{equation}\label{eq_Z}
   \frac{1}{Z}\frac{d^2 Z}{d z^2}-\frac{\omega^2}{c^2}\varepsilon_2 \cos(2 \beta z)=-k_z^2.
 \end{equation}
$k_z$ is a variable separation constant that does not depend on any coordinate variables. Eq.(\ref{eq_Z}) is the Mathieu equation \cite{MathieuEquation}, which has a general solution
\begin{equation}\label{Mathieu_cs}
  Z(z)=A_1\, C(\zeta_a,\zeta_q,\beta z)+ A_2\, S(\zeta_a,\zeta_q,\beta z),
\end{equation}
where $C(\zeta_a,\zeta_q,\beta z)$ and $S(\zeta_a,\zeta_q,\beta z)$ are the Mathieu cosine and sine functions respectively, $\zeta_a\equiv k_z^2/\beta^2$, $\zeta_q\equiv\varepsilon_2\omega^2 /(2 c^2 \beta^2)$, and $A_{1,2}$ are constants.

Now we are ready to obtain the general solution of $E_z$. Inserting Eq.(\ref{eq_Z}) into Eq.(\ref{RTZ_1}), the $R(r)$ function satisfies the Bessel equation and can be written as $R(r)\sim J_m(\gamma r)$, where $J_m$ is the Bessel function of the first kind and $\gamma^2=\omega^2(1+\varepsilon_1)/c^2-k_z^2$, for $r<a$.
A similar procedure can be done for $r>a$, with $R(r)\sim K_m(\tilde{\gamma}r)$, and $K_m$ is the modified Bessel function of the second kind and $\tilde{\gamma}^2=k_z^2-\omega^2/c^2$.

Having now the expression of $E_z$, we can insert it into the Maxwell equations to calculate the transverse component $E_r$ and $E_\phi$ \cite{JacksonBook}.
Since the amplitude of the Bragg modulation $\varepsilon_2$ is assumed to be weak, as a zeroth order approximation $\varepsilon_2\rightarrow0$, Eq.~(\ref{eq_Z}) reduces to $\frac{1}{Z}\frac{d^2 Z}{d z^2}=-k_z^2$ whose solution is $Z(z)=e^{\pm i k_z z}$, giving the usual homogeneous waveguide result.
In this case, we obtain the well known HE/EH mode $k_z\sim\omega$ dispersion relation
\cite{JacksonBook,KienPRA17}:
\begin{equation}\label{HEEH_dispersion}
  \begin{split}
    &\left[ \frac{J_m^\prime (\gamma a)}{\gamma a J_m(\gamma a)}+\frac{K_m^\prime(\tilde{\gamma} a)}{\tilde{\gamma}a K_m(\tilde{\gamma}a)}       \right] \left[ \frac{n_1^2 J_m^\prime (\gamma a)}{\gamma a J_m(\gamma a)}+\frac{ n_2^2 K_m^\prime(\tilde{\gamma} a)}{\tilde{\gamma} a K_m(\tilde{\gamma}a)}       \right]\\
    &=\frac{m^2 k_z^2 c^2}{\omega^2} \left( \frac{1}{\gamma^2 a^2}+\frac{1}{\tilde{\gamma}^2 a^2}  \right)^2,
  \end{split}
\end{equation}
where $n_1=\sqrt{1+\varepsilon_1}$ and $n_2=1$ are the refractive indices inside and outside the waveguide, respectively.

In the homogeneous case, $k_z$ bears two roles simultaneously: first, as a variable separation constant; and second, as the Floquet exponent. Once the Bragg modulation is introduced, the degeneracy of those two roles breaks down so that $k_z$ is no longer a good quantum number representing correctly the momentum of the light field. Hence, the $k_z\sim\omega$ relation is not the proper dispersion relation for $\varepsilon_2 \neq 0$.
In order to calculate the correct dispersion relation, we need to apply the Floquet theory explicitly.

According to the Floquet's theorem, the solutions of the Mathieu equation Eq.~(\ref{Mathieu_cs}) can be written in the form \cite{MathieuEquation}
\begin{equation}\label{floquet_expression}
  Z(z)=e^{i\tilde{k}\,\beta z}f(\beta z)\quad \text{and}\quad \tilde{k}=\tilde{k}(k_z,\omega)
\end{equation}
where $\tilde{k}$ is the Mathieu characteristic exponent and $f(\beta z)$ is a periodic function with the period $\pi$ \cite{FloquetTheory}. In the form of Eq.(\ref{floquet_expression}), we can see that the correct momentum quantum number is
$\tilde{k}$ rather than $k_z$, and the $\tilde{k}\sim\omega$ relation is the true dispersion relation for the Bragg modulation.

Therefore, we need to eliminate the intermediate parameter $k_z$ in order to obtain the $k_z\sim\omega$ dispersion.
As an example, let's consider the case $a\rightarrow\infty$ where the system reduces to an 1D photonic crystal (with planar slabs). For the TE/TM mode, the $k_z\sim\omega$ relation is given by $k_z=\omega \sqrt{1+\varepsilon_1}/c$ \cite{JoannopoulosBook08}, which is a linear relation and does not contain any discontinuity. Meanwhile, by using Eq.~(\ref{floquet_expression}), $k_z$ can be inversely expressed as $k_z=g(\tilde{k},\omega)$, where $g(\tilde{k},\omega)$ is an analytic function defined by series expansions \cite{MathieuFunction}. By matching two expressions of $k_z$, we have
\begin{equation}\label{TETMdispersion}
  \frac{\omega \sqrt{1+\varepsilon_1}}{c}=g(\tilde{k},\omega).
\end{equation}
This is an implicit expression of the desirable $\tilde{k}\sim\omega$ dispersion relation.
Fig.~\ref{Waveguide_illustration}(d) shows the numerically calculated dispersion curve by using Eq.~(\ref{TETMdispersion}). This semi-analytical result matches nicely the existing
results obtained from \emph{ab initio} calculations in photonic crystal literature \cite{JoannopoulosBook08}.

Now we consider Bragg modulations inside a waveguide with a finite radius $a$.
For a given waveguide mode with angular momentum $m$ and a given order of the solution, e.g. HE$_{31}$ mode, to obtain the $\tilde{k}\sim \omega$ dispersion relation the calculation procedure would be similar: first we calculate its $k_z\sim\omega$ relation by using Eq.~(\ref{HEEH_dispersion}) \cite{KienPRA17}, and then eliminate the $k_z$ dependency by using Eq.~(\ref{TETMdispersion}). Besides the existence of a
cut-off frequency for certain modes \cite{JacksonBook}, the resulting dispersion curve will possess the band gap structure introduced by the Bragg modulation.
Note that the transversely confined periodic structure can support bound states in the continuum \cite{BulgakovPRA17}, however, since those modes lie above the light line, they are outside the scope of our consideration.

\emph{SOI corrections.} Next we consider the effect of the SOI on the band gap. Here we only consider fiber modes with paraxial light where its spin and the intrinsic orbital angular momentum are separable \cite{VitulloPRL17,EnkEurLett94}. In this case, the angular momentum is parallel to the propagation direction and is represented by the operator $\hat{l}_z=-i\partial_\phi$ whose eigenvalue is $m$.
Recent experiments \cite{VitulloPRL17} have demonstrated that for a homogeneous cylindrical waveguide, the transverse confinement will result in a SOI correction term for the transverse electric field $\mathbf{E}_t$ (see the Supplemental Material):
\begin{equation}\label{H0_Et}
   \left[ \nabla_t^2 +\frac{\omega^2 \varepsilon_r(r) }{c^2} \right]\mathbf{E}_t+ \hat{H}_{SO}\mathbf{E}_t=k_z^2 \mathbf{E}_t,
\end{equation}
where $\nabla_t^2$ is the transverse Laplacian, $\hat{H}_{SO}$ is the effective SOI interaction \cite{LearyPRA14}:
\begin{equation}\label{HSO}
  \hat{H}_{SO}=\frac{\delta(r-a)\Delta}{4\, k_z \, a^2}\left(\frac{1}{a}\partial_r -\frac{a}{r} \hat{s}_z \hat{l}_z  \right),
\end{equation}
where $\Delta=\sqrt{\varepsilon_1+1}-1$ is the dielectric jump on the waveguide boundary; $\hat{s}_z$ and $\hat{l}_z$ are the spin and orbital angular momentum operators respectively. They are both defined against the $z$ axis of the laboratory frame.

\begin{figure}
  \centering
  \includegraphics[width=8cm]{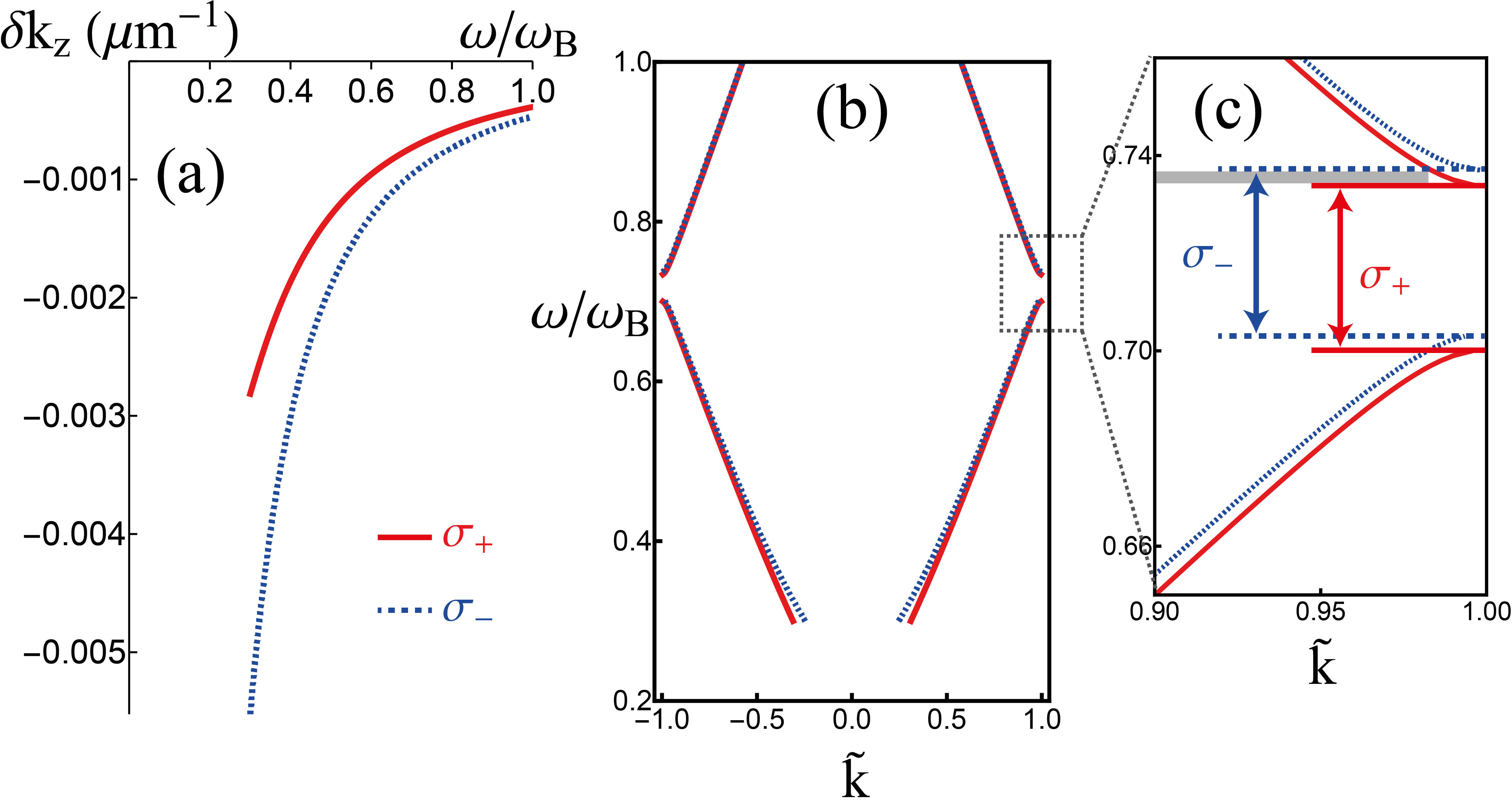}
  \caption{ SOI correction for the HE$_{31}$ mode. (a) Frequency dependent SOI corrections to $k_z$ for $\sigma_\pm$ components. (b) SOI corrected dispersion curves. (c) Zoom-in plot showing the shifted band gap between $\sigma_+$ and $\sigma_-$. Parameters are the same as in Fig.~\ref{Waveguide_illustration} with $a=5\, \text{$\mu$m}, \varepsilon_2=0.2, m=3, \text{and}\, \Delta=1.12$. In (b) and (c), dispersion curves are plotted in the helicity basis, and the splitting is magnified by a factor of $100$ for visualization purposes.}\label{fig_SOI_correction}
\end{figure}

One can see that Eq.~(\ref{H0_Et}) be seen as
$\mathbf{E}_t$ satisfying a vector Helmholtz equation with eigenvalues $k_z$, while $\hat{H}_{SO}$ is added as a perturbation. The perturbation against the eigenvalue can be conveniently calculated in the helicity basis
$\hat{\mathbf{e}}_\sigma=\frac{\hat{\mathbf{e}}_x+i \hat{\mathbf{e}}_y}{\sqrt{2}}\delta_{\sigma,+}+\frac{\hat{\mathbf{e}}_x-i \hat{\mathbf{e}}_y}{\sqrt{2}}\delta_{\sigma,-}$ \cite{LearyPRA14},
where $\sigma=\pm$ represents the right- and left- handedness of a photon's helicity and $\hat{\mathbf{e}}_{x,y}$ are the basis vectors of the Cartesian coordinate system. Up to the first order, assuming $m>1$ in the helicity basis, perturbations to $k_z$ read \cite{LearyPRA14}:
\begin{equation}\label{delta_kz}
\delta k_z^\sigma=\frac{\pi \Delta}{2\,k_z\,a^3}\int \delta(r-a) E_\sigma \left(r\frac{\partial}{\partial r}-\sigma\,m\right)E_\sigma \mathrm{d}r.
\end{equation}
The corrected eigenvalues $k_z^\sigma=k_z^0+\delta k_z^\sigma$ (where $k_z^0$ is the unperturbed eigenvalue) signify a splitting between the $\sigma_+$ and $\sigma_-$ components, which has been observed in experiments as the rotation of the spatial intensity pattern given by the interference of two beams with opposite orbital angular momentum \cite{LearyPRA14,VitulloPRL17}.
Fig.~\ref{fig_SOI_correction}(a) shows the dependence of the SOI correction $\delta k_z^\sigma$ on $\omega$ for the HE$_{31}$ mode away from the cut-off frequency, where the absolute value of $\delta k_z^\sigma$ is about $0.1\%$ of the original eigenvalue $k_z^0$.

Now we combine the SOI correction Eq.~(\ref{delta_kz}) with the Bragg modulation.
The splitting in $k_z$ gives rise to the splitting of the Floquet characteristic exponent $\tilde{k}$. If the zeroth order $k_z^0\sim\omega$ relation of the HE$_{31}$ mode is denoted as $k_z^0=h_{31}(\omega)$, then according to Eq.~(\ref{floquet_expression}), the shifted dispersion is given by
\begin{equation}\label{shift_k_tilde}
  h_{31}(\omega)+\delta k_z^\sigma (k_z^0,m)=g(\tilde{k},\omega).
\end{equation}
Therefore, the photonic band gap will be shifted correspondingly. Fig.~\ref{fig_SOI_correction}(b)(c) shows the calculated split band gap structure. Represented by the grey area, the gap shift under current parameters is about $3\times 10^{-3}\,\omega/\omega_B$, which is about $0.04$ THz.

The split band gap between helicity $\sigma_+$ and $\sigma_-$ leads to a single propagation channel within a certain range of frequency. For example, as it is illustrated in Fig.~\ref{Waveguide_illustration}(b), when a Laguerre-Gauss beam couples to the waveguide and generates the targeted waveguide mode \cite{VitulloPRL17}, in the shaded frequency area in Fig.~\ref{fig_SOI_correction}(c) only $\sigma_+$ can propagate, while $\sigma_-$ will be reflected.
When we transfer back to the laboratory basis and observe the spin (denoted by $s_\pm$), since the spin flips upon reflection by the gap, the final result is the spin-direction-locked propagation similar to that of CQO configurations. But it does not require any atom coupling to achieve chiral response.

To achieve the single-photon spin splitter, a single photon emitter/coupler \cite{AharonovichNatPho16,PetersenScience14} can be placed in the middle of the waveguide [Fig.~\ref{Waveguide_illustration}(c)]. Within a suitable frequency range, detectors on each side of the waveguide will receive spin-locked signals.
Compared to macrosopic polarizing filters, the merit of our structure is that it is based on the intrinsic SOI and can work in the nano-scale which is a crucial prerequisite in the field of integrated photonics.

\begin{figure}
  \centering
  \includegraphics[width=8cm]{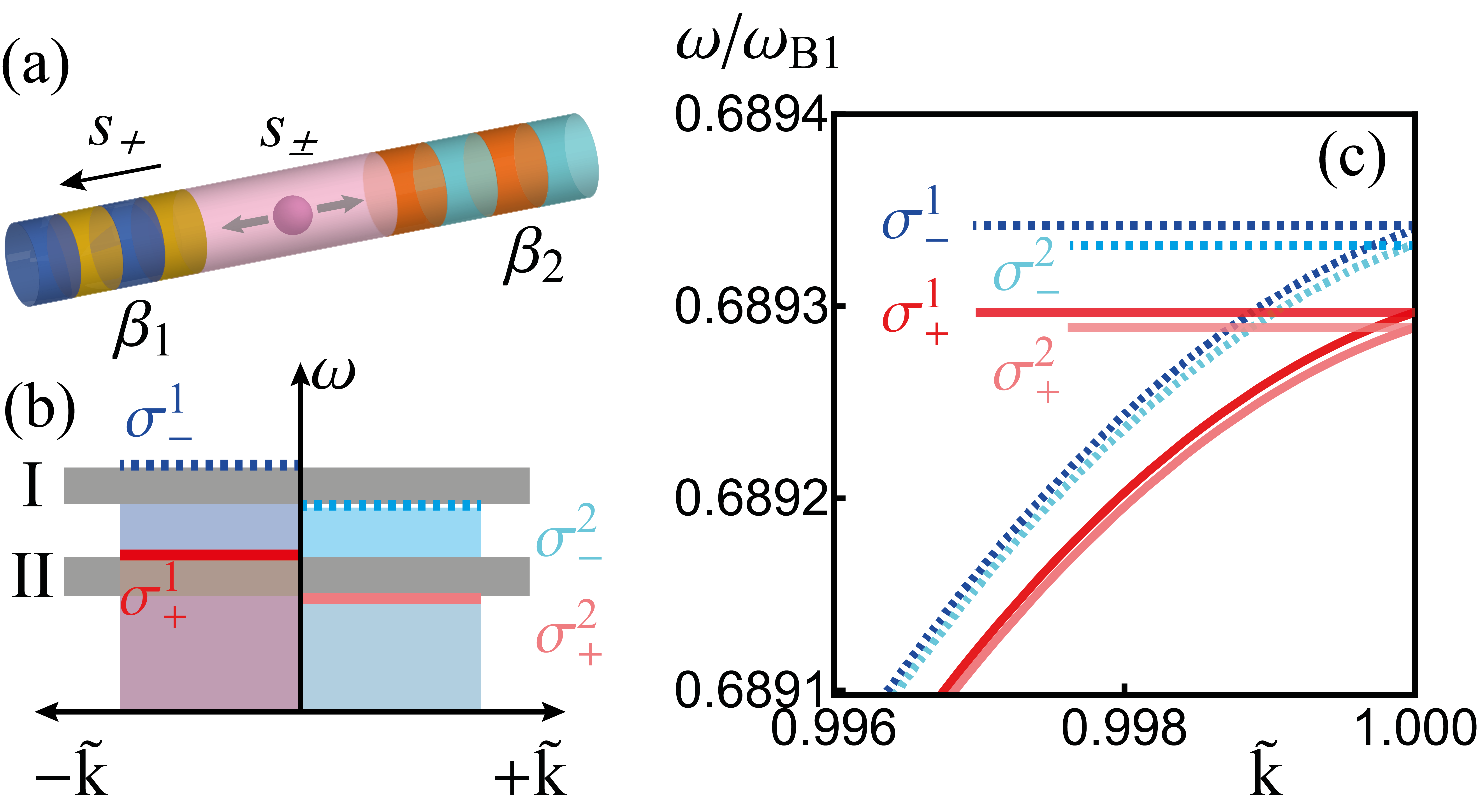}
  \caption{(a) An illustration of a single photon emitter/coupler embedded in the middle two Bragg modulation sections with difference spatial frequency.
   (b) An illustration of the lower boundary of the split band gap. The shaded areas are the corresponding dielectric bands. (c) The calculated lower boundary of a band gap for the HE$_{31}$ mode. Parameters are the same in Fig.~\ref{fig_SOI_correction} with $\omega_{B1}=\omega_B, \beta_1=\beta, \omega_{B2}=0.99999\, \omega_{B1}, \text{and}\, \beta_2=\omega_{B2}/c$. No magnification of splitting is included.}\label{two_blocks}
\end{figure}

Further, the single photon emitter can be placed between two blocks of Bragg modulations with slightly different spatial modulation frequency or amplitude, as illustrated in Fig.~\ref{two_blocks}(a). In this case, the band gaps for $+\tilde{k}$ and $-\tilde{k}$ will be different (calculated in helicity basis). Fig.~\ref{two_blocks}(c) shows an example calculation for the lower side of a band gap, where curves with darker colors are for left block (toward $-\tilde{k}$) and curves with lighter colors are for the right block (toward $+\tilde{k}$). In both blocks, $\sigma_\pm$ are split by the SOI effect so that there are four shifted band gaps to consider. A simplified illustration is shown in Fig.~\ref{two_blocks}(b), where the lower boundary of a band gap is shown by solid or dashed lines, and the shaded areas are the corresponding dielectric bands where light can propagate through \cite{JoannopoulosBook08}.

Figure~\ref{two_blocks}(a) shows the effect when light is emitted within the frequency region \textrm{I} listed in Fig.~\ref{two_blocks}(b). In this frequency range, only $\sigma_-$ can propagate toward $-\tilde{k}$ direction. Seen in the laboratory basis, only the $s_+$ can be detected on the left side of the waveguide, while other polarization components will be trapped within the effective microcavity formed by two blocks of Bragg modulations. Similarly, in frequency region \textrm{II} in Fig.~\ref{two_blocks}(b), only $s_+$ propagating toward the positive $z$ direction will be blocked, while all other polarization components can propagate freely.
This single-channel photon emission/blockage design can be implemented in quantum computation, spectroscopy and metrology, where directional emission of photons are vital \cite{LounisRPP05,PolyakovJMO09}.

In summary, we proposed a fully optical design for a single-photon spin splitter based on the SOI of light originating from the transverse confinement of a cylindrical waveguide. The SOI leads to a splitting between the propagation constant $k_z$ between the helicity $\sigma_+$ and $\sigma_-$ components. When a Bragg modulation of the dielectric constant is included, one needs to eliminate $k_z$ to obtain the genuine $\tilde{k}\sim\omega$ dispersion relation, where $\tilde{k}$ is the Floquet exponent. The SOI splitting in $k_z$ results in splitting in the dispersion and the splitting of the photonic band gap between two helicity components. When viewed in the laboratory basis, the split band gap provides a spin-locked propagation channel for two polarization states $s_+$ or $s_-$, forming the single-photon spin splitter.

The calculation procedure we proposed can be used to investigate the unidirectional invisibility \cite{LinPRL11} resulting from a parity ($\mathcal{P}$) and time ($\mathcal{T}$) symmetric dielectric distribution $\varepsilon_2 \cos(2\beta z)+i \varepsilon_2 \sin(2\beta z)=\varepsilon_2 \exp(i 2 \beta z)$. In this case, the $Z(z)$ function in Eq.~(\ref{eq_Z}) allows analytic solutions, hence its dispersion between the Floquet exponents and frequency can be calculated explicitly (see Supplementary Material). By studying the properties of the Floquet exponents, we might obtain insight into the spectral singularities \cite{AliPRL09,AliPRA11} resulting from the complex $\mathcal{PT}$-symmetric potentials.

\begin{acknowledgements}
G.L. acknowledges the EPSRC Programme on Hybrid Polaritonics for financial support and thanks Dr Ben Hopkins for useful discussions.
AK acknowledges the support from the
Russian Foundation for Basic Research (RFBR) and Deutsche Forschungsgemeinschaft (DFG) in the framework of International Collaborative Research Center TRR 160 (Project No. 15-52-12018).
The work was carried out with financial
support from the Ministry of Education and Science of the Russian Federation in the framework of increase Competitiveness
Program of NUST ”MISIS”, implemented by a governmental decree dated 16th of March 2013, No 211.
\end{acknowledgements}

\end{document}